\documentstyle[aps,prl,multicol,epsf]{revtex}
\begin{document}

\def\datestamp{March 15, 2004}

\def\bra#1{\left\langle {#1} \right|}
\def\ket#1{\left| {#1} \right\rangle}

\def\horparallel{
\epsfysize=2ex
\lower.5ex\hbox{
\epsffile{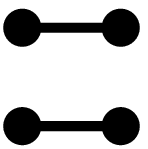}}\,\, }

\def\vertparallel{
\epsfysize=2ex
\lower.5ex\hbox{
\epsffile{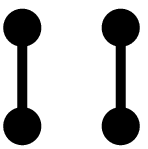}}\,\, }

\title{
Vison gap in the Rokhsar--Kivelson dimer model on the
triangular lattice}

\author{D.~A.~Ivanov}
\address{
Institute for Theoretical Physics, EPFL, CH-1015 Lausanne, Switzerland}

\date\datestamp
\maketitle

\begin{abstract}
With the classical Monte Carlo method, I find the energy gap in the
Rokhsar--Kivelson dimer model on the triangular lattice. I identify the
lowest excitations as visons, and compute their energy as a function
of the momentum.
\end{abstract}

\begin{multicols}{2}

The Rokhsar--Kivelson (RK) dimer model is an interesting example of a
two-dimensional quantum liquid \cite{Rokhsar-Kivelson,%
Moessner-Sondhi,Moessner-Sondhi-Fradkin}.
When the kinetic and potential coupling constants of the RK dimer model
are equal (the so-called ``RK point''), the ground state of the model is 
known exactly, and the ground-state properties on 
planar lattices may be analytically
studied with the Pfaffian method \cite{Kasteleyn}. 
Much attention was paid to the
study of the RK model on the triangular lattice, in which case the RK point
is known to be in the liquid phase with topological order and with
exponentially decaying correlations \cite{Moessner-Sondhi,Fendley,IIF} 
(on the square lattice, in contrast,
the RK point has power-law correlations and a gapless spectrum, 
due to the bipartite nature of the lattice \cite{Rokhsar-Kivelson,%
Moessner-Sondhi-Fradkin,Fisher-Stephenson,Levitov,Henley}).
Unfortunately, our knowledge of the excitation spectrum at the RK point
on the triangular lattice is limited to numerical methods 
\cite{Moessner-Sondhi,Ioffe}.
Studying excitations in the RK model is interesting in view
of the prediction that the elementary excitations are the 
so-called ``visons'' which
are non-local objects in terms of dimers. Visons were proposed as elementary
excitations in quantum liquids with $Z_2$ gauge symmetry  
(possibly also in spin-1/2 systems)\cite{Read-Chakraborty,Senthil-Fisher}. 
While applicability
of this proposal to realistic spin-1/2 systems is debatable, there exist
model systems, where the existence of vison excitations is explicitly
shown and the whole spectrum of visons may
be exactly found \cite{Misguich-Serban-Pasquier,Ioffe-Feigelman}. 
The usual price to pay for the exact solvability is that
in such models the vison excitations are strictly local (i.e., their
energy is independent of the momentum) and non-interacting. 
In contrast, the RK model
presents a very non-trivial case, where the visons have their dispersion
and interact in many-vison states. 

In this work, I use the proposal of Henley to extract the energy gap of the
excitations from the time correlation of the classical Monte Carlo algorithm
used to compute the ground-state observables \cite{Henley} 
(the classical calculation
of the excitation spectrum is actually possible due to the supersymmetry of the
RK point \cite{SUSY}; the relation between supersymmetric 
quantum systems and stochastic classical dynamics was discussed in various
contexts \cite{SUSY-2}). By choosing appropriate
correlation functions, we can resolve the momentum-dependence of the gap,
and also distinguish excitations carrying odd number of visons from those
carrying even number of visons. We find that the low-energy excitations
are indeed vison-like (carry odd number of visons) and plot their dispersion
in the Brillouin zone.

The Rokhsar--Kivelson dimer model may be defined on any graph: the
dimer coverings of the graph define an orthonormal basis of the 
Hilbert space; and the quantum Hamiltonian is
\begin{equation}
H_{RK}=\sum \Big( -t \left| \horparallel  \right\rangle\left\langle 
\vertparallel \right|
+ v
\left| \horparallel  \right\rangle\left\langle \horparallel \right|
\Big)
\label{RK-Hamiltonian}
\end{equation}
The sum is taken over all length-four loops of the graph.
The two coupling constants $t$ and $v$ determine the strength of
kinetic and potential terms, respectively. At the ``RK point'' $t=v$,
the Hamiltonian can be shown to be non-negative (assuming $t>0$), and
its ground state may be constructed as the sum of all dimer configurations
taken with equal amplitudes (more precisely, we can restrict the sum to
the dimer configurations from any connected component of the ``phase space'',
i.e. to configurations which can be obtained from each other by the
kinetic term in the Hamiltonian (\ref{RK-Hamiltonian}))
\cite{Rokhsar-Kivelson}. This ground state
has energy zero, which is a manifestation of the supersymmetry of the
RK point \cite{SUSY}. We further specify to the case of the underlying graph
being the two-dimensional triangular lattice (the sum in the Hamiltonian
(\ref{RK-Hamiltonian}) is taken over all rhombi, so the total number of 
terms in the sum is three times the number of lattice sites), and set
the energy units $t=v=1$. Our main objective in this paper
is finding the low-lying excitations of this model. From earlier numerical
studies, it has been suggested that this model has a gapped spectrum 
\cite{Moessner-Sondhi,Ioffe},
which seems to be in agreement with the exponential decay of ground-state 
correlation functions known from analytic studies \cite{Moessner-Sondhi,%
Fendley,IIF}.

\begin{figure}
\epsfxsize=1.0\hsize
\centerline{\epsfbox{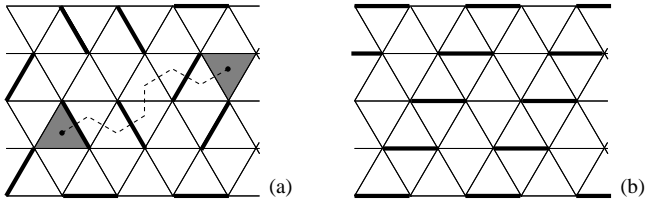}}
\medskip
\narrowtext
\caption{
{\bf (a)} Definition of the vison operator $V_\Gamma$. The contour $\Gamma$
(dashed line) connects two triangular plaquets of the lattice (shaded).
The value of $V_\Gamma$ is the parity of the intersection of dimers with
$\Gamma$. For the contour $\Gamma$ and for the dimer configuration shown in
the figure, $V_\Gamma=-1$ (three intersections). {\bf (b)} The reference
dimer configuration for the sign fixing of a single-vison operator.
An alternative (to that described in the main text) formulation
of the sign-fixing rule: the contours $\Gamma$ must be drawn in 
such a way that they do not intersect dimers from the reference configuration.
}
\label{fig-1}
\end{figure}

From the general discussions of the dimer liquids in two dimensions,
the $Z_2$-vortex operator (the so called ``vison'') is known to play an
important role among physical observables. 
The vison (more precisely, the ``two-vison'') operator $V_\Gamma$ is
defined for any contour $\Gamma$ intersecting links of the lattice 
and terminating
either at the lattice boundary or inside a plaquet in the bulk of 
the lattice (Fig.~\ref{fig-1}a). 
The operator $V_\Gamma$ is defined as the parity
of the number of dimers intersecting $\Gamma$:
\begin{equation}
V_\Gamma= (-1)^{{\rm no.~of~dimers~intersecting~}\Gamma}.
\label{vison-def}
\end{equation}
If one commutes such an operator with the Hamiltonian
(\ref{RK-Hamiltonian}), the only non-vanishing contribution comes from the
rhombi containing the end points of the contour. In particular, if the
contour forms a closed loop or terminates at a lattice boundary, the
corresponding operator exactly commutes with the Hamiltonian and gives rise
to different topological sectors of the Hilbert space \cite{Rokhsar-Kivelson,%
Read-Chakraborty,Bonesteel}. 

It is natural to suggest that terminating the contour in the bulk of the
lattice produces an excited state close to an eigenstate. Of course, to
form a true excited eigenstate would require some ``dressing'' of the vison
operator with local corrections around the contour end point. However,
the topological structure of the excitation will be preserved vison-like.
To clarify the above reasoning, we first describe some properties of the
vison operators $V_\Gamma$. First, it is easy to check that the operator
$V_\Gamma$, up to a sign, depends only on the end points of the contour; 
the dependence on the contour itself reduces to a controllable change of
the sign: changing from the contour $\Gamma$ to another contour $\Gamma'$
with the same end points changes the sign of the vison operator as
$V_\Gamma=(-1)^S V_{\Gamma'}$, where $S$ is the number of lattice points
between the contours $\Gamma$ and $\Gamma'$. Second, concatenation of the
contours corresponds to the multiplication of the corresponding vison
operators. Therefore we may represent the operator $V_\Gamma$ as
the product of two vison operators at end points: $V_\Gamma=V_1 V_2 $, 
where each of the ``single-vison''
operators $V_1$ and $V_2$ depends on one of the two end points of $\Gamma$. 
Constructed this way, the point vison operators $V_i$ obey $Z_2$ algebra
($V_i^2=1$) and are defined on the {\em frustrated dual lattice}: their
index $i$ refers to a plaquet of the original lattice, and they change
sign on going around one lattice site of the original lattice.
For the triangular lattice, we should think of visons as living on the
hexagonal lattice with the magnetic flux of half quantum per hexagon.

To establish a sign convention for visons, we need to fix a $Z_2$ gauge on 
the dual lattice. This can be most easily done by
taking a certain (arbitrary, but fixed once forever)  dimer configuration as
a reference one. Then in the definition (\ref{vison-def}), 
we multiply the right-hand side by the same expression $V_\Gamma$ 
computed in the 
reference dimer configuration. With the new definition, $V_\Gamma$ becomes
a single-valued function of the end points of $\Gamma$ (independent of the
choice of the contour).
In our calculation, we take the reference
dimer configuration as shown in Fig.~\ref{fig-1}b (note an additional doubling 
of the unit cell).

\begin{figure}
\epsfxsize=0.7\hsize
\centerline{\epsfbox{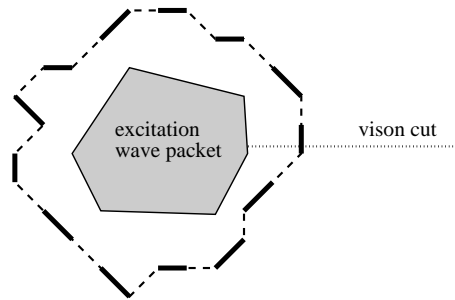}}
\medskip
\narrowtext
\caption{
A localized wave packet of vison-like excitations. Vison-like excitations
may be distinguished from non-vison-like ones by a circular
permutation of dimers along a big contour (shown). If we denote such a 
permutation by $P$, vison-like excitations obey the equation $P^2=-P$,
while non-vison-like excitations have $P^2=P$.
}
\label{fig-2}
\end{figure}

An important property of the vison operator is that it is a non-local
operator in terms of dimers. A single-vison operator $V_i$ in an infinite
system involves the contour $\Gamma$ continued to infinity and hence
corresponds to a change in the boundary conditions on the wave function
at infinity. Namely, a circular permutation of dimers along a big closed
contour encircling the ``excitation region'' reverses sign of the 
excitation with odd number of visons, but keeps the wave function unchanged
for the excitation with even number of visons (Fig.~\ref{fig-2}). 
According to this
criterion, we may classify any ``localized'' wave packet as either vison-like
or non-vison-like. Non-vison-like excitations are expressed as local
operators in terms of dimers. A vison-like excitation may be described
as a non-vison-like (local) operator multiplied by a vison operator.
Thus we naturally have two classes of excitations with a $Z_2$ grading:
combining two vison-like excitations we obtain a non-vison like excitation,
while combining a non-vison-like excitation with a vison-like excitation
gives a vison-like excitation. Of course, with this construction it seems
natural that vison-like excitations should be considered as ``elementary''
excitations, while non-vison-like excitations may be constructed as
composite excitations from vison-like elementary excitations. In reality,
however, it may happen that vison-like excitations are pushed high in
energy, so that the low-lying physical excitations are all non-vison-like.
In this paper I demonstrate that it is not the case at the RK point
on the triangular lattice. We shall see that both the vison-like and
non-vison-like sectors are gapped and that the gap in the vison-like
sector is smaller than in the non-vison-like sector. So vison excitations
are indeed the lowest-energy excitations in the model.

The exponentially decaying ground-state correlation functions 
\cite{Moessner-Sondhi,Fendley}
suggest a gap in the excitation spectrum, and indeed both quantum
Monte Carlo studies \cite{Moessner-Sondhi} and exact 
diagonalization on small systems \cite{Ioffe}
suggest the presence of the gap (from exact diagonalization, the value of
the gap was estimated as 0.1). However, as pointed out by Henley \cite{Henley},
at the RK point the gap may be more easily extracted from a classical
Monte Carlo simulation similar to that used for calculating the ground-state
expectation values (see, e.g., Refs.~\onlinecite{Balents-Fisher-Girvin,IIF}). 
Namely, consider the following
random walk defined on the space of all dimer coverings of the lattice.
A step of the random walk is defined as picking at random any rhombus
and, if it contains two parallel dimers,  flipping this pair of
dimers into the other two sides of the rhombus (as indicated by the
kinetic term of the Hamiltonian (\ref{RK-Hamiltonian})). If the chosen
rhombus is non-flippable, the dimer configuration remains unchanged at this
step of the random walk. 

As shown by Henley \cite{Henley}, such a random walk 
simulates the quantum-mechanical
evolution in imaginary time. Accordingly, the exponents governing the
decay of the dynamic correlation functions with time are given precisely
by the excitation energies of the quantum system. For our discrete random
walk, this procedure of determining the excitation energies produces
systematic errors arising from the discretization of time steps.
The discreteness of time steps may, in principle, be properly compensated;
however, we simply neglect the corresponding systematic errors.
One can show that discretization of time leads to relative 
corrections to the gap magnitude of approximately one over the total
number of rhombi in the system. In our Monte Carlo calculation we take
a sufficiently large system of 20$\times$20 sites (thus containing 1200
rhombi), and those corrections are smaller than the statistical errors
for the lengths of random walks used in our calculations. Therefore we
disregard the time-discretization errors and extract the gaps directly
from the correlations of the discrete random walk.

The next useful observation is that, by taking appropriate correlation
functions, we can probe the gap at a given wave vector. And, moreover,
we can distinguish between vison-like and non-vison-like excitation
sectors. For computing the gap in the non-vison-like sector, 
we should consider correlations
of non-vison-like observables (e.g., the dimer density). For
the gap in the vison sector we take vison-like observables 
(e.g., the point vison $V_i$ defined above).

We first consider the excitations in the {\it vison-like} sector. In order
to compute the gap, we take the correlation function
\begin{equation}
F({\bf r}_{ij}, t-t') = \langle V_i(t) V_j(t') \rangle\, ,
\label{vison-corr}
\end{equation}
where $V_i(t)$ is the point vison on the triangle $i$ at the
moment $t$ of the random-walk procedure as defined above, ${\bf r}_{ij}$
is the vector connecting the plaquets $i$ and $j$. 
This correlation function is properly defined, in
spite of having only one vison operator at each of the time moments
$t$ and $t'$. To demonstrate the consistency of the definition 
(\ref{vison-corr}),
it is convenient to insert the square of the vison operator $[V_j(t)]^2=1$.
Then we rewrite $V_i(t) V_j(t')=(V_i(t)V_j(t))(V_j(t)V_j(t'))$. The first
product of the two vison operators is defined in (\ref{vison-def}) with
the contour $\Gamma$ connecting the points $i$ and $j$. The second product
involves two vison operators at one space point, but at different time
moments. Obviously this product is also well defined for our random-walk
process: namely, every dimer flip at any time between $t$ and $t'$ 
on a rhombus containing the triangle $j$ changes the sign of $V_j(t)V_j(t')$.
In other words, $V_j(t)V_j(t')$ counts the parity of the number of such
flips.

\begin{figure}
\epsfxsize=0.7\hsize
\centerline{\epsfbox{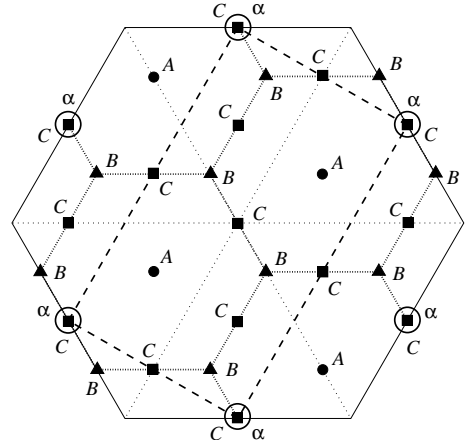}}
\medskip
\narrowtext
\caption{
This figure shows the Brillouin zones for non-vison-like 
and vison-like excitations (superimposed, on the same scale).
{\bf Non-vison-like excitations:} the Brillouin zone is the big
hexagon (solid line; the side length of the hexagon is 4$\pi$/3
in the units of the inverse lattice constant of the original 
triangular lattice), 
with the minimum-gap points marked with big
circles and labeled $\alpha$. {\bf Vison-like excitations:} the
Brillouin zone is the dashed rectangle (for the choice of gauge
specified in Fig.~\ref{fig-1}b). The small solid circles, squares,
and triangles (labeled with the letters $A$, $B$, and $C$, respectively)
are the high-symmetry points. For example, points $A$ are centers
of six-fold symmetry. The vison gap is found to reach its minimum 
at points $B$, and to increase strongly towards points $A$.
}
\label{fig-3}
\end{figure}

With the gauge choice for visons as discussed above (Fig.~\ref{fig-1}b), 
the unit
cell of the lattice is doubled, and contains four triangular plaquets.
To characterize vison excitations with wave vectors, we perform a
Fourier transform of the correlation function (\ref{vison-corr}) and
arrive at the 4$\times$4 matrix $F({\bf k},t)$, where ${\bf k}$ is the
wave vector from the reduced Brillouin zone. The full Brillouin zone
and the reduced Brillouin zone are sketched in Fig.~\ref{fig-3}. 
Note that with
our choice of the vison gauge, the correlation function $F({\bf k},t)$
has certain symmetries in the $k$-space reflecting the symmetries of the
original triangular lattice. Those symmetries contain the six-fold
rotation/reflection symmetry ($D_6$ symmetry group) around a certain
point in the $k$-space, together with translations by two basis vectors
defining a triangular lattice. The high-symmetry points in the $k$-space
are marked in Fig.~\ref{fig-3} by letters $A$, $B$, and $C$.

\begin{figure}
\centerline{\epsfxsize=0.8\hsize
\epsfbox{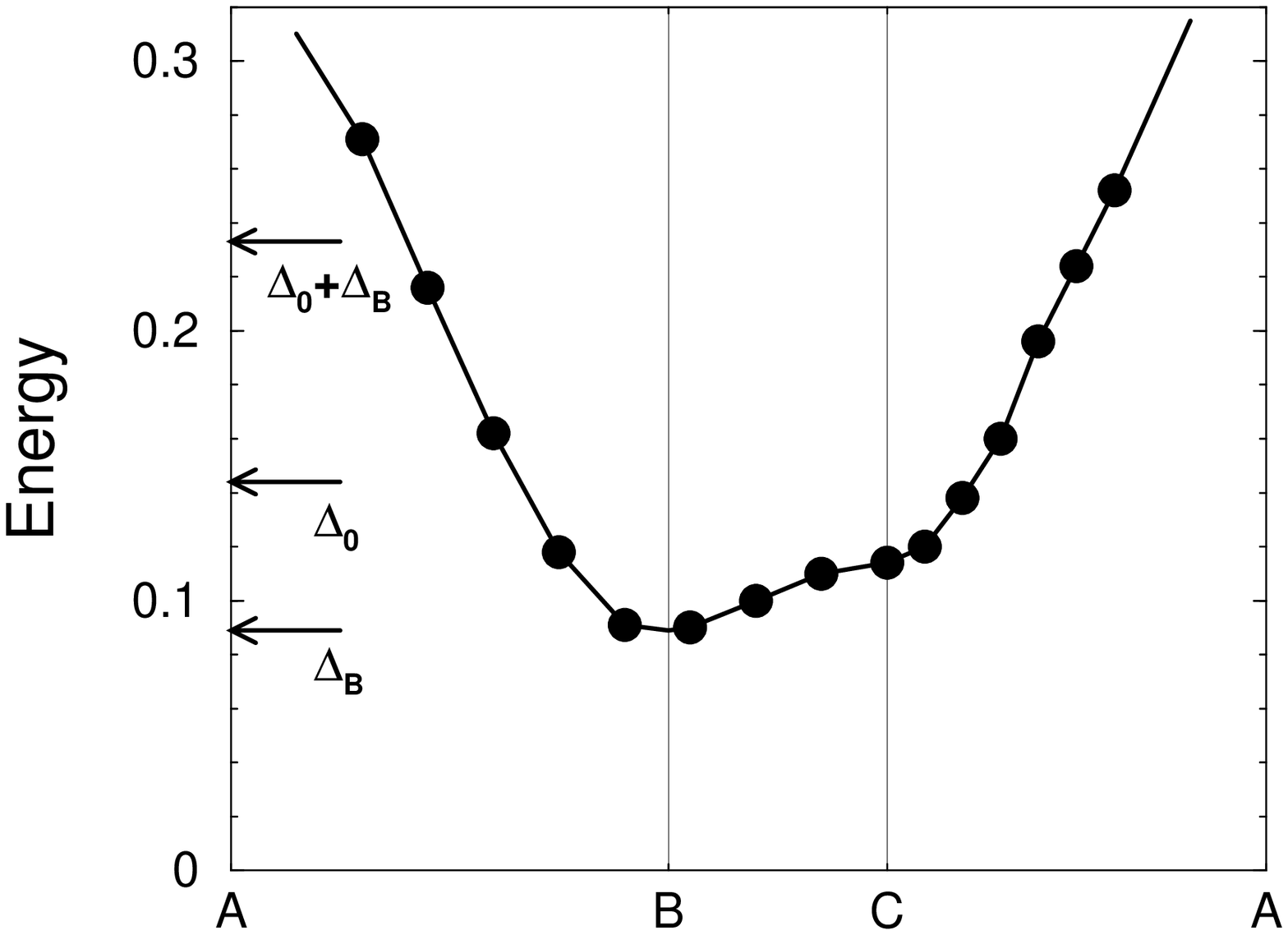}} 
\medskip
\centerline{\epsfxsize=0.8\hsize
\epsfbox{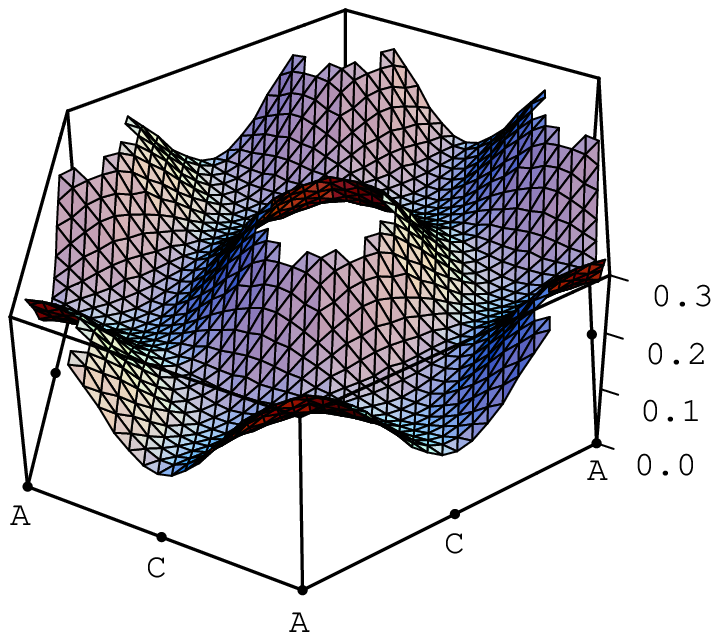}}
\medskip
\narrowtext
\caption{{\bf Top:} the energy of the vison-like excitations along
the section $A$--$B$--$C$--$A$ of the Brillouin zone shown in 
Fig.~\protect\ref{fig-3}. The error bars are much smaller than the
symbol size.
The arrows on the energy axis indicate
positions of $\Delta_B$ (bottom of the vison band), $\Delta_0$ 
(lowest-energy non-vison-like excitation), and $\Delta_B + \Delta_0$
(an estimate for the bottom of the three-vison continuum).
{\bf Bottom:} three-dimensional plot of the energy of the vison-like
excitations as a function of the wave vector. 
The center and the corners of the hexagon correspond to the points $A$,
as labeled in Fig.~\protect\ref{fig-3}.
}
\label{fig-4}
\end{figure}

\begin{figure}
\centerline{
\epsfxsize=0.8\hsize
\epsfbox{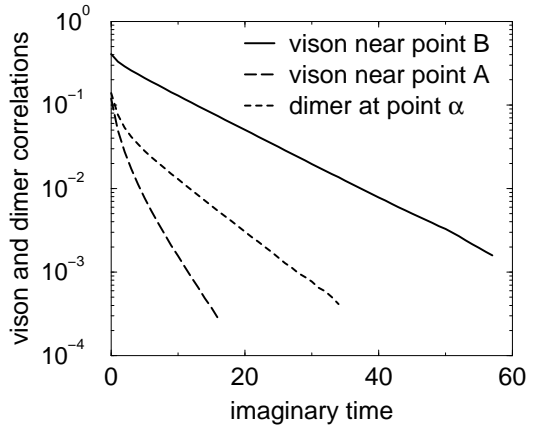}}
\medskip
\narrowtext
\caption{
Typical vison and dimer correlations in the classical Monte Carlo
random walk. Solid an dashed line show the vison correlations for
wave vectors near points $B$ and $A$, respectively 
(see Fig.~\protect\ref{fig-3}). Note that while near point $B$ the
correlation function has a well-marked exponential behavior at long
times, the correlations near point $A$ do not allow a precise fitting with
a single exponent. The dotted line shows the dimer-dimer correlation
function at the point $\alpha$ (the minimum-gap point). A good exponential
decay of the correlation function possibly indicates 
a two-vison bound state.
}
\label{fig-5}
\end{figure}

Now we determine the energy of the vison excitations $\Delta({\bf k})$
from the exponential
decay of the diagonal elements of $F({\bf k},t)$ \cite{detail}. 
The sections of the
energy dispersion along the ``crystallographic axes'' of the Brillouin
zone and the 3-D plot of $\Delta({\bf k})$ are presented in Fig.~\ref{fig-4}.
The points $B$ of the $k$-space give the minimal energy gap
$\Delta_B=0.089(1)$ [which agrees with Ref.~\onlinecite{Ioffe} claiming 
the gap value 0.1]. The points $C$ are the saddle points of the
energy dispersion with $\Delta_C=0.114(1)$. The points $A$ are the
centers of high-energy regions. 
In those regions, a naive fitting with an exponential suggests 
$\Delta_A>0.3$ and does not reproduce well the $t$-dependence 
of $F({\bf k},t)$. This may indicate that, at those wave vectors, the lowest
excitations are not elementary visons, but rather combinations of three
visons from the points $B$. Thus the $t$-dependence of $F({\bf k},t)$
reflects not an isolated excitation, but the bottom of a multi-particle
continuum. The difference in the time dependence between
low-energy and high-energy regions in the $k$ space is illustrated in
Fig.~\ref{fig-5} where we show typical $t$-dependences of $F({\bf k},t)$ for
the two $k$-points: one near point $B$ in the $k$-space, and the other one near
point $A$.

Next, we repeat the same procedure for {\it non-vison-like}
excitations by taking the correlations
of the dimer-density operator instead of the vison operator $V_i(t)$ 
in (\ref{vison-corr}). We then find that the lowest gap in the non-vison
sector is $\Delta_0=0.144(1)$ and is reached at points labeled $\alpha$
in Fig.~\ref{fig-3}. The $t$-dependence of the dimer-dimer correlation function
at point $\alpha$ is shown in Fig.~\ref{fig-5}. 
A good fit with an exponential
dependence, together with the inequality  $\Delta_0< 2 \Delta_B$, suggests
that the lowest non-vison-like excitation is not just a superposition of two
non-interacting vison excitations, but a bound state of such a pair.
On the other hand, its energy is considerably higher than that of an
elementary vison excitation ($\Delta_B$), which confirms the claim that
the lowest excitation is vison-like. [This result also suggests to estimate
the bottom of the continuum in the vison-like excitations as 
$\Delta_0+\Delta_B$ instead of $3\Delta_B$, see Fig.~\ref{fig-4}a.]

Finally, with this method of calculating the excitation gap, we can verify
the claim of Ref.\onlinecite{Ioffe} about the absence of low-lying edge states
in the case of lattices with boundaries. We examine the excitation spectrum
of the 10$\times$10 and 20$\times$20 cylinders with straight boundaries along 
the lattice directions. From the result on non-vison-like excitations 
in the bulk, we
expect that visons are attracted to each other, and therefore, visons should 
also get attracted to boundaries (since the vison cut may be terminated at the
boundary at no energy cost). Note also that near the boundary there is no
distinction between vison-like and non-vison-like excitations. Thus for
determining the energy of the edge states we may take the
dimer-dimer correlation
function at the very boundary. Our classical Monte Carlo simulation 
gives the boundary gap
$\Delta_{\rm edge}=0.072(1)$ reached at the wave vector $\pi$ along 
the boundary. In line with our expectations, the gap at the boundary is indeed
somewhat reduced compared to the bulk vison gap, but remains finite.

The author thanks M.~Feigelman for helpful discussions and comments.

\end{multicols}
\end{document}